\documentclass[12pt]{article}
\usepackage{amsmath}
\usepackage{amssymb}
\usepackage{cite}
\usepackage{epsfig}

\psfull

\catcode`\%=12
\catcode`\|=14
\newcommand\apa[3]    {\@spires{APASA
        {{\it Acta Phys.\ Austriaca }{\bf #1} (#2) #3}}
\newcommand\apas[3]    {\@spires{APAUA
        {{\it Acta Phys.\ Austriaca, Suppl.\ }{\bf #1} (#2) #3}}
\newcommand\appol[3] {\@spires{APPOA
        {{\it Acta Phys.\ Polon.\ }{\bf #1} (#2) #3}}
\newcommand\advm[3]  {\@spires{ADMTA
        {{\it Adv.\ Math.\ }{\bf #1} (#2) #3}}
\newcommand\adnp[3]   {\@spires{ANUPB
        {{\it Adv.\ Nucl.\ Phys.\ }{\bf #1} (#2) #3}}
\newcommand\adp[3]   {\@spires{ADPHA
        {{\it Adv.\ Phys.\ }{\bf #1} (#2) #3}}
\newcommand\atmp[3] {\@spires{00203
        {{\it Adv.\ Theor.\ Math.\ Phys.\ }{\bf #1} (#2) #3}}
\newcommand\am[3]    {\@spires{ANMAA
        {{\it Ann.\ Math.\ }{\bf #1} (#2) #3}}
\newcommand\ap[3]    {\@spires{APNYA
        {{\it Ann.\ Phys.\ (NY) }{\bf #1} (#2) #3}}
\newcommand\araa[3] {\@spires{ARAAA
        {{\it Ann.\ Rev.\ Astron.\ \& Astrophys.\ }{\bf #1} (#2) #3}}
\newcommand\arnps[3] {\@spires{ARNUA
        {{\it Ann.\ Rev.\ Nucl.\ Part.\ Sci.\ }{\bf #1} (#2) #3}}
\newcommand\asas[3]   {\@spires{AAEJA
        {{\it Astron.\ Astrophys.\ }{\bf #1} (#2) #3}}
\newcommand\asj[3]   {\@spires{ANJOA
        {{\it Astron.\ J.\ }{\bf #1} (#2) #3}}
\newcommand\app[3]   {\@spires{APHYE
        {{\it Astropart.\ Phys.\ }{\bf #1} (#2) #3}}
\newcommand\apj[3]    {\@spires{ASJOA
        {{\it Astrophys.\ J. }{\bf #1} (#2) #3}}
\newcommand\baas[3]   {\@spires{AASBA
        {{\it Bull.\ Am.\ Astron.\ Soc.\ }{\bf #1} (#2) #3}}
\newcommand\bams[3]   {\@spires{BAMOA
        {{\it Bull.\ Am.\ Math.\ Soc.\ }{\bf #1} (#2) #3}}
\newcommand\blms[3]   {\@spires{LMSBB
        {{\it Bull.\ London Math.\ Soc.\ }{\bf #1} (#2) #3}}
\newcommand\cjm[3]  {\@spires{CJMAA
        {{\it Can.\ J.\ Math.\ }{\bf #1} (#2) #3}}
\newcommand\cqg[3]   {\@spires{CQGRD
        {{\it Class.\ and Quant.\ Grav.\ }{\bf #1} (#2) #3}}
\newcommand\cmp[3]   {\@spires{CMPHA
        {{\it Commun.\ Math.\ Phys.\ }{\bf #1} (#2) #3}}
\newcommand\ctp[3]   {\@spires{CTPMD
        {{\it Commun.\ Theor.\ Phys.\ }{\bf #1} (#2) #3}}
\newcommand\cag[3]   {\@spires{00142
        {{\it Commun.\ Anal.\ Geom.\ }{\bf #1} (#2) #3}}
\newcommand\cpam[3]   {\@spires{CPAMA
        {{\it Commun.\ Pure Appl.\ Math.\ }{\bf #1} (#2) #3}}
\newcommand\cpc[3]   {\@spires{CPHCB
        {{\it Comput.\ Phys.\ Commun.\ }{\bf #1} (#2) #3}}
\newcommand\dmj[3]   {\@spires{DUMJA
        {{\it Duke Math.\ J. }{\bf #1} (#2) #3}}
\newcommand\epjc[3]  {{\it Eur.\ Phys.\ J. }{\bf C #1} (#2) #3}
\newcommand\epjd[3]  {\@spires{EPHJD
        {{\it Eur.\ Phys.\ J. Direct.\ }{\bf C #1} (#2) #3}}
\newcommand\epl[3]    {\@spires{EULEE
        {{\it Europhys.\ Lett. }{\bf #1} (#2) #3}}
\newcommand\forp[3]    {\@spires{FPYKA
        {{\it Fortschr.\ Phys.\ }{\bf #1} (#2) #3}}
\newcommand\faa[3]    {\@spires{FAAPB
        {{\it Funct.\ Anal.\ Appl.\ }{\bf #1} (#2) #3}}
\newcommand\grg[3]    {\@spires{GRGVA
        {{\it Gen.\ Rel.\ Grav.\ }{\bf #1} (#2) #3}}
\newcommand\hpa[3]   {\@spires{HPACA
        {{\it Helv.\ Phys.\ Acta }{\bf #1} (#2) #3}}
\newcommand\ijmpa[3] {\@spires{IMPAE
        {{\it Int.\ J.\ Mod.\ Phys.\ }{\bf A #1} (#2) #3}}
\newcommand\ijmpb[3] {\@spires{IMPAE
        {{\it Int.\ J.\ Mod.\ Phys.\ }{\bf B #1} (#2) #3}}
\newcommand\ijmpc[3] {\@spires{IMPAE
        {{\it Int.\ J.\ Mod.\ Phys.\ }{\bf C #1} (#2) #3}}
\newcommand\ijmpd[3] {\@spires{IMPAE
        {{\it Int.\ J.\ Mod.\ Phys.\ }{\bf D #1} (#2) #3}}
\newcommand\ijtp[3] {\@spires{IJTPB
        {{\it Int.\ J.\ Theor.\ Phys.\ }{\bf #1} (#2) #3}}
\newcommand\invm[3]  {\@spires{INVMB
        {{\it Invent.\ Math.\ }{\bf #1} (#2) #3}}
\newcommand\jag[3]   {\@spires{00124
        {{\it J.\ Alg.\ Geom.\ }{\bf #1} (#2) #3}}
\newcommand\jams[3]   {\@spires{00052
        {{\it J.\ Am.\ Math.\ Soc.\ }{\bf #1} (#2) #3}}
\newcommand\jap[3]   {\@spires{JAPIA
        {{\it J.\ Appl.\ Phys.\ }{\bf #1} (#2) #3}}
\newcommand\jdg[3]   {\@spires{JDGEA
        {{\it J.\ Diff.\ Geom.\ }{\bf #1} (#2) #3}}
\newcommand\jgp[3]   {\@spires{JGPHE
        {{\it J.\ Geom.\ Phys.\ }{\bf #1} (#2) #3}}
\newcommand\jhep[3]  {{\it J. High Energy Phys.\ }{\bf #1} (#2) #3}
\newcommand\jmp[3]   {\@spires{JMAPA
        {{\it J.\ Math.\ Phys.\ }{\bf #1} (#2) #3}}
\newcommand\joth[3]  {\@spires{JOTHE
        {{\it J.\ Operator Theory }{\bf #1} (#2) #3}}
\newcommand\jpha[3]   {\@spires{JPAGB
        {{\it J. Phys.\ }{\bf A #1} (#2) #3}}
\newcommand\jphc[3]   {\@spires{JPAGB
        {{\it J. Phys.\ }{\bf C #1} (#2) #3}}
\newcommand\jphg[3]   {\@spires{JPAGB
        {{\it J. Phys.\ }{\bf G #1} (#2) #3}}
\newcommand\lmp[3]   {\@spires{LMPHD
        {{\it Lett.\ Math.\ Phys.\ }{\bf #1} (#2) #3}}
\newcommand\ncl[3]    {\@spires{NCLTA
        {{\it Lett.\ Nuovo Cim.\ }{\bf #1} (#2) #3}}
\newcommand\matan[3]  {\@spires{MAANA
        {{\it Math.\ Ann.\ }{\bf #1} (#2) #3}}
\newcommand\mussr[3]  {\@spires{MUSIA
        {{\it Math.\ USSR Izv.\ }{\bf #1} (#2) #3}}
\newcommand\mams[3]  {\@spires{MAMCA
        {{\it Mem.\ Am.\ Math.\ Soc.\ }{\bf #1} (#2) #3}}
\newcommand\mpla[3]  {\@spires{MPLAE
        {{\it Mod.\ Phys.\ Lett.\ }{\bf A #1} (#2) #3}}
\newcommand\mplb[3]  {\@spires{MPLAE
        {{\it Mod.\ Phys.\ Lett.\ }{\bf B #1} (#2) #3}}
\newcommand\nature[3]  {\@spires{NATUA
        {{\it Nature }{\bf #1} (#2) #3}}
\newcommand\nim[3]   {\@spires{NUIMA
        {{\it Nucl.\ Instrum.\ Meth.\ }{\bf #1} (#2) #3}}
\newcommand\npa[3]   {\@spires{NUPHA
        {{\it Nucl.\ Phys.\ }{\bf A #1} (#2) #3}}
\newcommand\npb[3] {{\it Nucl.\ Phys.\ }{\bf B #1} (#2) #3}
\newcommand\npps[3]  {\@spires{NUPHZ
        {{\it Nucl.\ Phys.\ }{\bf #1} {\it(Proc.\ Suppl.)} (#2) #3}}
\newcommand\nc[3]    {\@spires{NUCIA
        {{\it Nuovo Cim.\ }{\bf #1} (#2) #3}}
\newcommand\ncs[3]  {\@spires{NUCUA
        {{\it Nuovo Cim.\ Suppl.\ }{\bf #1} (#2) #3}}
\newcommand\pan[3]  {\@spires{PANUE
        {{\it Phys.\ Atom.\ Nucl.\ }{\bf #1} (#2) #3}}
\newcommand\pla[3]   {\@spires{PHLTA
        {{\it Phys.\ Lett.\ }{\bf A #1} (#2) #3}}
\newcommand\plb[3]   {{\it Phys.\ Lett.\ }{\bf B #1} (#2) #3}
\newcommand\pr[3]    {\@spires{PHRVA
        {{\it Phys.\ Rev.\ }{\bf #1} (#2) #3}}
\newcommand\pra[3]   {\@spires{PHRVA
        {{\it Phys.\ Rev.\ }{\bf A #1} (#2) #3}}
\newcommand\prb[3]   {\@spires{PHRVA
        {{\it Phys.\ Rev.\ }{\bf B #1} (#2) #3}}
\newcommand\prc[3]   {\@spires{PHRVA
        {{\it Phys.\ Rev.\ }{\bf C #1} (#2) #3}}
\newcommand\prd[3] {{\it Phys.\ Rev.\ }{\bf D #1} (#2) #3}
\newcommand\pre[3]   {\@spires{PHRVA
        {{\it Phys.\ Rev.\ }{\bf E #1} (#2) #3}}
\newcommand\prep[3]  {\@spires{PRPLC
        {{\it Phys.\ Rept.\ }{\bf #1} (#2) #3}}
\newcommand\prl[3] {{\it Phys.\ Rev.\ Lett.\ }{\bf #1} (#2) #3}
\newcommand\phys[3]   {\@spires{PHYSA
        {{\it Physica }{\bf #1} (#2) #3}}
\newcommand\plms[3]   {\@spires{PHLTA
        {{\it Proc.\ London Math.\ Soc.\ }{\bf B #1} (#2) #3}}
\newcommand\pnas[3]  {\@spires{PNASA
        {{\it Proc.\ Nat.\ Acad.\ Sci.\ }{\bf #1} (#2) #3}}
\newcommand\ppnp[3]  {\@spires{PPNPD
        {{\it Prog.\ Part.\ Nucl.\ Phys.\ }{\bf #1} (#2) #3}}
\newcommand\ptp[3]   {\@spires{PTPKA
        {{\it Prog.\ Theor.\ Phys.\ }{\bf #1} (#2) #3}}
\newcommand\ptps[3]   {\@spires{PTPSA
        {{\it Prog.\ Theor.\ Phys.\ Suppl.\ }{\bf #1} (#2) #3}}
\newcommand\rmp[3]   {\@spires{RMPHA
        {{\it Rev.\ Mod.\ Phys.\ }{\bf #1} (#2) #3}}
\newcommand\sjnp[3]  {\@spires{SJNCA
        {{\it Sov.\ J.\ Nucl.\ Phys.\ }{\bf #1} (#2) #3}}
\newcommand\sjpn[3]  {\@spires{SJPNA
        {{\it Sov.\ J.\ Part.\ Nucl.\ }{\bf #1} (#2) #3}}
\newcommand\jetp[3]  {\@spires{SPHJA
        {{\it Sov.\ Phys.\ JETP\/ }{\bf #1} (#2) #3}}
\newcommand\jetpl[3]  {\@spires{JTPLA
        {{\it Sov.\ Phys.\ JETP Lett.\ }{\bf #1} (#2) #3}}
\newcommand\spu[3]  {\@spires{SOPUA
        {{\it Sov.\ Phys.\ Usp.\ }{\bf #1} (#2) #3}}
\newcommand\tmf[3]   {\@spires{TMFZA
        {{\it Teor.\ Mat.\ Fiz.\ }{\bf #1} (#2) #3}}
\newcommand\tmp[3]   {\@spires{TMPHA
        {{\it Theor.\ Math.\ Phys.\ }{\bf #1} (#2) #3}}
\newcommand\ufn[3]   {\@spires{UFNAA
        {{\it Usp.\ Fiz.\ Nauk.\ }{\bf #1} (#2) #3}}
| }}}}}}}}}}}}}}}}}}}}}} "|" is here a comment (catcode defined above) to
| }}}}}}}}}}}}}}}}}}}}}} include parenthesis for emacs to parse properly.
\newcommand\ujp[3]   {\@spires{00267
        {{\it Ukr.\ J.\ Phys.\ }{\bf #1} (#2) #3}}
\newcommand\yf[3]    {\@spires{YAFIA
        {{\it Yad.\ Fiz.\ }{\bf #1} (#2) #3}}
\newcommand\zpc[3]   {\@spires{ZEPYA
        {{\it Z.\ Physik }{\bf C #1} (#2) #3}}
\newcommand\zetf[3]  {\@spires{ZETFA
        {{\it Zh.\ Eksp.\ Teor.\ Fiz.\ }{\bf #1} (#2) #3}}

\newcommand{\newjournal}[5]{\@spires{#2
        {{\it #1 }{\bf #3} (#4) #5}}

\newcommand\ibid[3]{{\it ibid.\ }{\bf #1} (#2) #3}
\catcode`\%=14
\catcode`\|=12
\newcommand{\hepth}[1]{\href{http://xxx.lanl.gov/abs/hep-th/#1}{\tt hep-th/#1}}
\newcommand{\hepph}[1]{\href{http://xxx.lanl.gov/abs/hep-ph/#1}{\tt hep-ph/#1}}
\newcommand{\heplat}[1]{\href{http://xxx.lanl.gov/abs/hep-lat/#1}{\tt hep-lat/#1}}
\newcommand{\hepex}[1]{\href{http://xxx.lanl.gov/abs/hep-ex/#1}{\tt hep-ex/#1}}
\newcommand{\nuclth}[1]{\href{http://xxx.lanl.gov/abs/nucl-th/#1}{\tt nucl-th/#1}}
\newcommand{\nuclex}[1]{\href{http://xxx.lanl.gov/abs/nucl-ex/#1}{\tt nucl-ex/#1}}
\newcommand{\grqc}[1]{\href{http://xxx.lanl.gov/abs/gr-qc/#1}{\tt gr-qc/#1}}
\newcommand{\qalg}[1]{\href{http://xxx.lanl.gov/abs/q-alg/#1}{\tt q-alg/#1}}
\newcommand{\accphys}[1]{\href{http://xxx.lanl.gov/abs/accphys/#1}{\tt accphys/#1}}
\newcommand{\alggeom}[1]{\href{http://xxx.lanl.gov/abs/alg-geom/#1}{\tt alg-geom/#1}}
\newcommand{\astroph}[1]{\href{http://xxx.lanl.gov/abs/astro-ph/#1}{\tt astro-ph/#1}}
\newcommand{\chaodyn}[1]{\href{http://xxx.lanl.gov/abs/chao-dyn/#1}{\tt chao-dyn/#1}}
\newcommand{\condmat}[1]{\href{http://xxx.lanl.gov/abs/cond-mat/#1}{\tt cond-mat/#1}}
\newcommand{\nlinsys}[1]{\href{http://xxx.lanl.gov/abs/nlin-sys/#1}{\tt nlin-sys/#1}}
\newcommand{\quantph}[1]{\href{http://xxx.lanl.gov/abs/quant-ph/#1}{\tt quant-ph/#1}}
\newcommand{\solvint}[1]{\href{http://xxx.lanl.gov/abs/solv-int/#1}{\tt solv-int/#1}}
\newcommand{\suprcon}[1]{\href{http://xxx.lanl.gov/abs/supr-con/#1}{\tt supr-con/#1}}
\newcommand{\Math}[2]{\href{http://xxx.lanl.gov/abs/math.#1/#2}{\tt math.#1/#2}}
\newcommand{\arXivid}[1]{\href{http://arxiv.org/abs/#1}{\tt arXiv:#1}}
\def\order#1{{\cal{O}}\left(#1\right)}
\def\wrt{with respect to }

\def\vkti{\vec{k}_{ti}}
\def\vkt{\vec{k}_{t}}
\def\vpt{\vec{p}_{t}}
\def\pt{{p}_{t}}
\def\vb{\vec{b}}
\def\vp{\vec{p}}
\def\vn{\vec{n}}

\def\tkt{\kappa_t}
\def\tkti{\kappa_{t}}
\def\kt{k_t}
\def\kti{k_{ti}}
\def\ktj{k_{tj}}

\def\vtkt{\vec{\kappa}_t}
\def\vtkti#1{\vec{\kappa}_{t#1}}

\def\tpt{p}
\def\vtpt{\vec{\tpt}}

\def\da{\partial_a}

\def\HR{{\cal H}_{\mathrm{R}}}
\def\HL{{\cal H}_{\mathrm{L}}}
\def\rc{{\cal R}_{\mathrm{C}}}
\def\rr{{\cal R}_{\mathrm{R}}}
\def\ec{{\cal E}_{\mathrm{C}}}
\def\elim{{\cal E}_{\mathrm{lim}}}

\def\vb{\vec{b}}

\def\SigmaTilde{{\widetilde \Sigma}}
\def\CTilde{{\widetilde C}}
\def\Qbar{{\bar Q}}
\def\bmu{\bar \mu}
\def\tC{\tilde{D}}
\def\nubar{{\bar \nu}}
\def\bbar{{\bar b}}
\def\qbar{{\bar q}}
\newcommand{\Ft}{{\widetilde F}}
\newcommand{\Pt}{{\widetilde P}}

\newcommand{\bq}{\mathbf{q}}
\newcommand{\bP}{\mathbf{P}}
\newcommand{\bC}{\mathbf{C}}

\newcommand{\heavy}{\mathrm{heavy}}
\newcommand{\light}{\mathrm{light}}
\newcommand{\rightH}{\mathrm{right}}
\newcommand{\GeV}{\mathrm{GeV}}

\def\al{\alpha}
\def\be{\beta}
\def\gam{\gamma}
\def\lam{\lambda}
\def\om{\omega}
\def\Om{\Omega}

\def\bal{{\bar \alpha}}
\def\bbe{{\bar \beta}}
\def\deta{\Delta \eta}

\def\cS{{\cal{S}}}    
\def\cC{{\cal{C}}}    
\def\cD{{\cal{D}}}    
\def\cV{{\cal{V}}}    

\newcommand{\tot}{\mathrm{tot}}

\def\cR{{\cal{R}}}               
\newcommand{\dsig}{{d^2\!\sigma}}

\def\half{\mbox{\small $\frac{1}{2}$}}

\def\MSbar{\overline{\mbox{\scriptsize MS}}}
\def\DIS{\mathrm{DIS}}
\def\CMW{\textsc{cmw}}

\def\CF{C_F}
\def\TR{T_R}
\def\CA{C_A}
\def\NC{N_C}
\def\nf{n_{\!f}}

\def\as{\alpha_{{\textsc{s}}}}
\def\gae{{\gamma_{\textsc{e}}}}
\def\asb{{\bar \alpha}_{{\textsc{s}}}}

\def\ascmw{\alpha_{\CMW}}
\def\LQCD{\Lambda_{\mbox{\scriptsize QCD}}}

\def\ee{e^+e^-}

\newcommand\epjcd[3]  {
                {{\it Eur.\ Phys.\ J. Direct }{\bf C #1} (#2) #3}}

\begin{document}

\begin{titlepage}
\begin{flushright}
{MAN/HEP/2008/6 \\Bicocca-FT-08-4}
\end{flushright}
\vspace*{\fill}
\begin{center}
{\Large \textsf{\textbf{Azimuthal decorrelations between QCD jets at
all orders}}}
\end{center}
\par \vskip 5mm
\begin{center}

{\large \textsf{  A.~Banfi}} \\
Universit\`a degli studi di Milano-Bicocca and INFN, Sezione di Milano-Bicocca, Italy.\\
{\large \textsf{ M.~Dasgupta}}\\
School of Physics and Astronomy, University of Manchester \\
Manchester M13 9PL, U.K.\\
{\large \textsf{Y.~Delenda}}\\
D\'epartment de Physique, Facult\'e des Sciences,\\
Universit\'e de Batna, Algeria.
\end{center}
\par \vskip 2mm
\begin{center} {\large \textsf{\textbf{Abstract}}}\end{center}
\begin{quote}A quantity that promises to reveal important information
on perturbative and non-perturbative QCD dynamics is the azimuthal
decorrelation between jets in different hard processes. In order to
access this information fixed-order NLO predictions need to be
supplemented by resummation of logarithmic terms which are large in
the region where the jets are nearly back-to-back in azimuth. In the
present letter we carry out this resummation to next-to--leading
logarithmic accuracy explaining the important role played by the
recombination scheme in general resummations for such jet
observables.
\end{quote}
\vspace*{\fill}

\end{titlepage}

\section{Introduction}

One of the most commonly measured jet observables in experimental
QCD studies is the azimuthal decorrelation $\Delta \phi$ between
hard final-state jets. When compared to theoretical estimates of the
same, this quantity is expected to provide valuable information both
on QCD parameters (strong coupling, pdfs) as well as dynamics in the
near back-to--back region sensitive to multiple soft and/or
collinear emissions and non-perturbative effects. To this end it has
thus been often examined in experimental QCD studies at HERA and the
Tevatron~\cite{hera1,hera2, Hansson,Tevatron}, used for the tuning
of parameters of Monte Carlo event generator models and to constrain
unintegrated parton distribution functions (updfs) in conjunction
with HERA data~\cite{Jungpapera,Jungpaperb}.

Various complementary approaches are possible to study the dijet
$\Delta \phi$ including NLO calculations~\cite{Nagy}, resummation of
logarithms arising from the back-to--back region $\Delta \phi
\approx\pi$ as well as non-perturbative effects important in the
same region and in the small-$x$ regime the inclusion of BFKL
effects or the use, as mentioned before, of unintegrated parton
distribution functions~\cite{Jungpapera,Jungpaperb}.  However no
attempt has been made at examining the use of a combination of these
approaches in order to obtain general predictions valid over all of
the dijet phase-space and widest possible range of $\Delta \phi$.

In ref.~\cite{Hansson}, for instance, a problem was noted in the
comparison of NLO QCD calculations to H1 data on the $\Delta \phi$
distribution which was apparently resolved by the use of
updfs~\cite{Jungpapera} suggesting the importance of QCD dynamics
and non-perturbative effects in the probed kinematical region. One
is thus led to wonder whether other approaches based on say
conventional resummation in the back-to--back region may also help
to ameliorate the problems with pure NLO results.  Such resummation
would not include BFKL effects and it would be interesting to see at
what $x$ values genuine small-$x$ effects are actually needed by the
data.

However there has not been much progress in the case of final-state
resummation for observables such as this which, in contrast to the
much studied event-shape variables, are crucially dependent on the
exact definition of final-state jets. These are typically quantities
which are constructed from ``aggregate'' \emph{jet} kinematic
variables (momenta, azimuth, rapidities) of which other examples are
dijet invariant masses and jet $p_t$ spectra. Here the jet momenta
are obtained from the particle (hadron) momenta after running an
algorithm and specifying a recombination scheme. The algorithm
dictates which particles end up in the jet and the recombination
scheme how the jet kinematic quantities are related to those of its
constituent hadrons. Resummation in these cases is a far more
delicate affair and there are only a few instances in the literature
of resummed predictions for such jet
observables~\cite{SterKid,Birdflower}.

One of the main complications that arises in such problems is that,
as we shall illustrate, one is typically studying observables that
are sensitive to energy flow outside well-defined jet regions which
potentially means that many such observables fall into the category
of non-global QCD observables~\cite{DassalNG1,DassalNG2}. Since it
was shown that the resummation of non-global observables is
substantially more complicated than that for ``global'' quantities
such as most event-shape variables and in any case restricted to the
large $N_c$ approximation, the most accurate theoretical predictions
can be obtained only for global observables. This appears to rule
out the possibility of complete next-to--leading logarithmic
estimates for many interesting jet observables including potentially
the azimuthal decorrelation we study here. As far as existing
predictions for jet observables are concerned, the issue of
non-global logarithms was not dealt with in ref.~\cite{SterKid}
(published prior to the discovery of non-global effects) where they
would arise in threshold resummation for one of the definitions
($M^2 =(p_1+p_2)^2$) of the dijet invariant mass studied there but
would be absent for the definition $M^2 = 2 p_1.p_2$.  Further we
should also mention here that the non-global component has been
incorrectly treated in ref.~\cite{Birdflower} where it is mentioned
that such effects will vanish with jet radius when in fact one
obtains a saturation in the small $R$ limit as was explicitly shown
for the case of jets in ref.~\cite{BanDasSymm}.

In the present letter we shall show an interplay between the
potential non-global nature of the observable and the exact
definition of the jet as provided by the choice of a recombination
scheme. This may be taken as an example of how carefully selecting
the definition of the observable and the jets one may be able to
render an exact NLL resummation possible, avoiding altogether the
non-global issue.  To be precise, here we point out that in a
certain experimentally popular recombination scheme (used to study
dijet azimuthal decorrelations at HERA) the observable at hand is in
fact global and hence one can resum up to next-to--leading
logarithms exactly. In a different recombination scheme (currently
used at the Tevatron) the observable is non-global. However for the
particular case of azimuthal decorrelations we point to recent
developments which indicate that non-global logarithms while
formally present in the latter scheme will be numerically
insignificant here and should not substantially impede
phenomenological investigations near the back-to--back region. We
should mention explicitly that we do not advocate here the general
use of one recombination scheme over another: a scheme that has good
features theoretically for one observable may not be so good for
another and hence ideally speaking an observable-by--observable
choice is optimal.

This letter is organised as follows. In the following section we
derive the dependence of the quantity $\Delta \phi$ on multiple soft
emissions in two different recombination schemes and hence
distinguish its global and non-global variants. In the subsequent
section we provide resummed results in impact parameter space for
the global variant for both DIS and hadron collisions while pointing
out that we have also resummed the non-global variant to sufficient
accuracy. We then present our numerical results for the inverse
transform from impact parameter space and hence for the azimuthal
decorrelation distribution.  Lastly we briefly discuss our results
mentioning the further developments needed in terms of matching to
fixed-order calculations as well as including non-perturbative
effects and point to work in progress in this regard.

\section{Recombination scheme, kinematics and globalness}

We wish to study the impact of two recombination schemes used to
construct the angle $\Delta \phi$ between the final-state jets in
dijet production. In the first scheme~\cite{fermilab} the jet
azimuthal angle $\phi_j$ is given by a $p_t$-weighted sum over its
hadronic constituents, $\phi_j = \sum_{i \in j} p_{t,i}
\phi_i/\sum_{i \in j}p_{t,i}$, while in the second scheme one
constructs the jet four-vector $p_j = \sum_{i \in j} p_i$, with the
sum running over hadrons in the jet, and then parameterises $p_j =
p_{t,j} \left( \cosh
  \eta_{j}, \cos \phi_j, \sin \phi_j, \sinh \eta_j \right)$ to obtain
the jet azimuth $\phi_j$. The first scheme is employed for instance
by the H1 collaboration at HERA (see ref.~\cite{hanthesis}) while to
our knowledge the latter ($E$-scheme) is currently preferred by the
Tevatron experiments.

Having defined the relevant schemes let us consider the final-state
kinematics. The final-state configuration that concerns us here is
one where the hard jets are nearly back-to--back in azimuth and
hence the system is close to the Born configuration for dijet
production. In this limit other than the hard dijet system one has
to consider the presence of any number of soft emitted quanta which
cause a small deviation from $|\phi_{j1}-\phi_{j2}| = \Delta \phi =
\pi $. The transverse momenta of final-state particles can then be
parameterised as below\footnote{Here one is looking at the
projections of particle momenta in the plane perpendicular to the
beam direction in hadron collisions or that perpendicular to the
$\gamma^{*} P$ axis in the DIS Breit or HCM frames.}:
\begin{eqnarray}
\vec{p}_{t,1} &=& p_{t,1}(1,0),\nonumber \\
\vec{p}_{t,2} &=& p_{t,2}(\cos(\pi-\epsilon),\sin(\pi-\epsilon)),\nonumber \\
&=& p_{t,2}(-\cos\epsilon,\sin\epsilon),\nonumber \\
\vec{k}_{t,i} &=& k_{t,i}(\cos\phi_i,\sin\phi_i),
\end{eqnarray}
where the hard final-state partons are labeled by $1$ and $2$ and the
soft gluons by the label $i$. For only soft emissions the hard partons
are nearly back-to--back and $|\epsilon | \ll 1$.

In the scheme involving the $p_t$-weighted sum we write the azimuth of
the leading jets as:
\begin{eqnarray}
  \phi_{j1}&=&\frac{\sum_{i\in j1}k_{t,i}\, \phi_i}{p_{t,1}+\sum_{i\in
      \mathrm{j1}}k_{t,i}} \approx \frac{\sum_{i\in j1}k_{t,i}\phi_i}{p_t},\nonumber \\
  \phi_{j2} &=& \frac{\sum_{i\in j2 }k_{t,i}\,
    \phi_i+p_{t,2}(\pi-\epsilon)}{p_{t,2}+\sum_{i\in
      \mathrm{j2}}k_{t,i}} \approx (\pi-\epsilon) + \frac{\sum_{i\in
      j2}k_{t,i}(\phi_i-\pi)}{p_t},
\end{eqnarray}
where to obtain results correct to first order in the small
quantities $k_{t,i}$ it suffices to set $p_{t,1} =p_{t,2}=p_t$ and
by momentum conservation it follows that $\epsilon=-\sum_{i}
k_{t,i}\sin\phi_i/p_t$, discarding all correction terms quadratic in
soft momenta, that do not affect our results.

Note that the azimuth of the reconstructed jet 1 has a small
deviation from $\phi=0$, whereas that for jet 2 has a small
deviation from $\phi=\pi$, due to the emission of soft gluons. Hence
the effects of soft emission on the azimuthal angle (as measured by
the deviation from $\Delta \phi = \pi$) are given by:
\begin{equation}
\label{eq:kin} \left|\pi-\Delta\phi \right| =
\left|\sum_i\frac{k_{t,i}}{p_t} \left(\sin\phi_i-\theta_{i1}
\phi_i-\theta_{i2}(\pi-\phi_i)\right)\right| +{\mathcal{O}}
\left(k_{t}^2 \right),
\end{equation}
where $\theta_{ij}=1$ if particle $i$ is clustered to jet $j$ and is
zero otherwise. The definition above implies that the observable in
question is global since it is sensitive to soft emissions in the
whole phase-space, both in and outside the jets, and the dependence
on soft emissions in either case is linear in $k_t$. This property
ensures that it is possible to resum the large-logarithms in the
back-to--back region to next-to--leading (single) logarithmic
accuracy without resorting to the large $N_c$ approximation needed
for non-global observables~\cite{DassalNG1,DassalNG2}.

Now turning to the $E$-scheme, to obtain the corresponding
dependence on soft emissions, we construct the four-momentum of a
jet as $p^{\mu}_{j} = \sum_{i \in j} p_i^{\mu} $, where the sum runs
over all partons/hadrons in the jet. Thus one obtains, in
particular, the transverse momentum vector of the jet and hence the
angle $\phi_{j}$ in the transverse plane as the inverse-tangent of
the ratio of components, $\tan\phi_{j} = p_{t,y}/p_{t,x}$. Employing
this procedure one obtains the following result for the azimuthal
angle between jets in the four-vector recombination scheme:
\begin{equation}
  \left|\pi-\Delta\phi \right| = \left|\sum_{i \notin {\mathrm{jets}}}
    \frac{k_{t,i}}{p_t} \sin\phi_i \right|+{\mathcal{O}} \left(k_t^2
  \right),
\end{equation}
where the sum extends over all soft particles not recombined with
the hard jets. This result is natural since if all particles were
combined into the hard jets then by momentum conservation the jets
would be back-to--back in the plane transverse to the beam (no net
transverse momentum). Hence with four-vector addition deviations
from $\Delta \phi = \pi$ are only caused by particle flow outside
the jet regions. Observables sensitive to soft emissions in such
delimited angular intervals are of the non-global
variety~\cite{DassalNG1,DassalNG2}, and hence in the $E$-scheme
definition of jets the azimuthal decorrelation is a non-global
observable.

\section{Resummation}

Having established that the observable at hand is a global
observable in the $p_t$-weighted recombination scheme its
resummation is now straightforward. It resembles closely resummation
for the azimuthal decorrelations between hadrons studied in
ref.~\cite{BanSmyeMark} as well as the resummation of dijet rates in
the region of symmetric $E_t$ cuts~\cite{BanDasSymm}. While
refereing the reader to the above references for more detailed
explanations of the steps involved, we sketch the main arguments
briefly below.

For the case of dijets produced in DIS one is examining soft radiation
off a three-hard-parton antenna (taking account of the incoming parton
in addition to the two hard partons that form final-state jets). One
may thus describe the probability for $n$ soft gluon emission by the
essentially classical form~\cite{yupinbanzan}:
\begin{equation}
\label{eq:fact} |M_n|^2 = |M_\mathcal{B}|^2 \frac{1}{n!}
\prod_{i=1}^{n}W(k_i),
\end{equation}
with $M_\mathcal{B}$ the matrix element for the process with no
emissions (Born order) and where one has the following gluon
emission probability in the eikonal approximation:
\begin{equation}
W(k_i) = g_s^2 \frac{N_c}{2}\left(w_{q_1g}(k_i)+ w_{q_2g}(k_i)
-\frac{1}{N_c^2} w_{q_1q_2}(k_i) \right),
\end{equation}
for emission of a soft gluon $k_i$ off a three-hard-parton system
comprising quarks $q_1$, $q_2$ and a gluon $g$. The $w_{ij}$
factors are just standard dipole antennae, with $w_{ij}(k) = \left(
  p_i.p_j \right)/\left((p_i.k)(p_j.k)\right)$ representing the
emission from the various dipoles formed by the three parton system,
and $g_s^2 = 4 \pi \alpha_s$ is the strong coupling\footnote{The
argument of the coupling is (to our next-to--leading logarithmic
(NLL) accuracy) the transverse momentum of $k_i$ with respect to the
dipole axis in the dipole rest frame $\kappa_{t,ij}^2 =
2/w_{ij}(k)$~\cite{yupinbanzan}.}.

Given the factorised nature of the eikonal squared matrix element
eq.~\eqref{eq:fact}, it is only needed to factorise the phase-space in
order to show exponentiation up to next-to--leading logarithmic
accuracy. The phase-space constraint for computing the cross-section
integrated up to some value of $\Delta \phi$ can be expressed as the
condition that one is considering events with $|\pi-\Delta \phi| <
\Delta$, where for small $\Delta$ one is in the back-to-back region.
One can then obtain the $\Delta$ distribution or equivalently the
distribution in $\Delta \phi$ by straightforward differentiation with
respect to $\Delta$. The integrated cross-section is (schematically)
given by
\begin{multline}
\sigma(\Delta) =  \sum_{a=q,g} \int d \mathcal{B}\>
|M^a_\mathcal{B}|^2\>
f_a(x,\mu_f^2)\sum_{n}\frac{1}{n!}\int \prod_{i=1}^n
[dk_i]\> W_a(k_i) \times \\ \times \Theta \left(\Delta -
\left|\sum_{i=1}^{n} \frac{k_{t,i}}{p_t} \left(\sin
\phi_i-\theta_{i1} \phi_i-\theta_{i2}(\pi-\phi_i)\right)\right|
\right),
\end{multline}
where we used a compact notation with the squared matrix element for
lowest-order dijet production given by $|M^a_\mathcal{B}|^2$, and we
integrate over the dijet configuration $\int\! d\mathcal{B}$
including the experimental cuts, as well as denote by $\int [dk_i]$
the integration over soft gluon momentum components. Further the
index $a$ represents the type of incoming parton (quark or gluon)
and $f_a(x,\mu_f^2)$ the parton density with $\mu_f$ a factorisation
scale\footnote{$x$ is not to be confused with Bjorken-$x$, it is the
momentum fraction carried by the incoming rather than the struck
parton and we have $x =x_B/\xi$ where $x_B$ is Bjorken-$x$ and $\xi
= Q^2/(2p.q)$ with $p$ the momentum of the incoming parton.}. In
brackets we have the step function constraint that restricts real
emission contributions while soft virtual emissions are
unconstrained and will be included later by imposing unitarity.
Since the emission probability for multiple soft gluon factorises as
indicated above, to achieve a resummed result it only remains to
factorise the phase-space condition by using a Fourier
representation of the step function:
\begin{equation}
\Theta \left( \Delta -\left|\sum_{i} v(k_i)\right| \right) =
\frac{1}{\pi} \int_{-\infty}^{\infty} \frac{db}{b} \sin \left(b
\Delta \right) \prod_i e^{ibv(k_i)}.
\end{equation}
Using the factorised matrix element and phase space allows us to
exponentiate the single gluon emission result in $b$-space and one
obtains:
\begin{equation}\label{eq:sigma-fin}
\sigma(\Delta) = \sum_{a=q,g} \int d\mathcal{B}\>
|M^a_{\mathcal{B}}|^2 \>\Sigma_a \left(\{ p \},\Delta \right),
\end{equation}
with
\begin{equation}
\Sigma_a (\{p \},\Delta) = \frac{1}{\pi}
\int_{-\infty}^{\infty}\frac{db}{b}  \sin(b \Delta)
e^{-R_a(b)}f_a(x,\mu_f^2).
\end{equation}
The function $R_a(b)$, known as the radiator, embodies the soft
single-gluon result which exponentiates in $b$-space. It contains a
characteristic dependence on the hard parton configuration that we
will presently explicate. This dependence is represented by the
dependence on the set of Born momenta $\{ p \} $ of the function
$\Sigma$ above.

Then we have for the radiator the standard-looking result:
\begin{equation}
  \label{eq:R-start}
  R_a(b) = \int \frac{d^3 k}{2 (2 \pi)^3 k_0} W_a(k) \left(1-\exp[ib
    v(k)]\right),
\end{equation}
where we also introduced virtual corrections via the unity in
parenthesis.  Noting that to NLL accuracy one can replace
$\left(1-\exp [i bv(k)] \right) \to 1-\cos(ib v(k)) \to \Theta
\left(v(k) -1/\bar{b}\right)$, where $\bar{b}=b e^{\gamma_E}$, we can
write:
\begin{equation}
\label{eq:intcross} \Sigma_a(\Delta,\{p\}) = \frac{2}{\pi}
\int_0^{\infty} \frac{db}{b}  \sin(b \Delta) \exp[-R_a(\bar{b})]
f_a\left(x,\mu_f^2/\bar{b}^2 \right),
\end{equation}
and carry out the computation for
$R_a(\bar{b})$,\footnote{$R(\bar{b})$ can be written in the
well-known exponentiated form~\cite{CTTW} $R(\bar{b}) =Lg_1(\alpha_s
L)+g_2(\alpha_s L)+\alpha_s(g_3 \alpha_s L)+\cdots$, where $L = \ln
\bar{b}$. NLL accuracy in our notation amounts to the complete
computation of the $g_1$ and $g_2$ functions.}

\begin{equation}
  R_a(\bar{b}) =\int \frac{d^3k}{2(2 \pi)^3k_0} W_a(k)
  \left(v(k)-\bar{b}^{-1}\right).
\end{equation}

We have thus far accounted only for soft emissions. To extend the
result to include hard collinear radiation we need to extend the
computation of the radiator such that in the collinear limit we use
the full QCD splitting functions instead of just the infrared pole
pieces contained in the $w_{ij}$ antenna functions. Moreover a set
of hard collinear emissions on the incoming leg are accommodated by
a change of scale in the parton distributions $f_a(x,\mu_f^2) \to
f_a(x,\mu_f^2/\bar{b}^2)$ via DGLAP evolution. Since this step is
standard and common to $b$-space resummations with incoming partons
we do not display its derivation, but for a fuller treatment we
point the reader to ref.~\cite{BanZanSmye}.

The result for $R_a(\bar{b})$ including the extension for hard
collinear radiation can be expressed in terms of three pieces each
with a distinct physical origin:
\begin{equation}
R_a(\bar{b}) =
R^a_{\mathrm{in}}(\bar{b})+R^a_{\mathrm{out}}(\bar{b}) -\ln
S\left(\bar{b},\{p\}\right),
\end{equation}
with $R^a_{\mathrm{in}}$ and $R^a_{\mathrm{out}}$ being the
contributions generated by emissions collinear to the incoming
(excluding the set of single-logarithms already resummed in the
parton densities) and outgoing legs respectively. In addition to
these jet functions we have a soft function $S(\bar{b},\{p\}) $
which resums soft emissions at large angles, and which depends on
the geometry of the emitting hard ensemble expressed here as a
dependence on the set of hard Born momenta $\{p\} $.

While our results eventually include the two-loop running of the
coupling which is necessary to obtain full NLL accuracy (compute the
full functions $g_1$ and $g_2$), for brevity and to illustrate the
main features we report our results here in a fixed coupling
approximation. In this case we simply obtain:
\begin{eqnarray}
\label{eq:raddis}
R^a_{\mathrm{out}}(\bar{b}) &=& (C^a_1+C^a_2) \frac{\alpha_s}{2 \pi}
\left(\frac{2}{3} L^2 + \frac{4}{3} L \left(-\ln 3 -4 \ln 2+3
\ln \frac{Q}{p_t} \right) \right)+\nonumber\\
&&+\frac{4}{3} \frac{\alpha_s}{2\pi}
\left(C^a_1 B^a_1+C^a_2 B^a_2\right) L ,\\
R^a_{\mathrm{in}}(\bar b) &=& C^a_i \frac{\alpha_s}{2 \pi} \left(2 L^2 +4L
\left(-\ln 2 + \ln \frac{Q}{p_t} \right) \right)
+ 4 C^a_i \frac{\alpha_s}{2\pi} B^a_i L, \\
\ln S(\bar{b},\{p\}) &=& - 4 L\left(2 C_F \frac{\alpha_s}{2 \pi} \ln
\frac{Q_{qq'}}{Q} + C_A \frac{\alpha_s}{2 \pi} \ln
\frac{Q_{qg}Q_{gq'}}{Q_{qq'}Q} \right),
\end{eqnarray}
with $L=\ln \bar{b}$. In the above $C^a_i$ is the colour charge of
the incoming parton in channel $a$, for instance $C^a_i=C_F$ for
$a=q$, the incoming quark channel. Likewise $C^a_{1,2}$ are the
colour charges of the partons initiating the outgoing jets $1$ and
$2$ in channel $a$. The main aspect of the results for the collinear
$R^a_{\mathrm{out,in}}$ jet functions is a leading double
logarithmic behaviour, where one notes the unfamiliar coefficient
$2/3$ (different from all commonly studied event-shape variables for
instance) associated to the double logs on the outgoing legs, i.e.
in the function $R^a_{\mathrm{out}}$. Additionally hard collinear
radiation is described by single-logarithmic terms with the
coefficients $C_\ell B_\ell$ for each leg, with the appropriate
colour charge $C_\ell$ ($\ell=i,1,2$) and $B_{i,1,2}$ depending on
the identities (spins) of the incoming and outgoing partons such
that $ B_\ell = -3/4$ for fermions and $B_\ell= -(11 C_A -4 T_R
n_f)/(12 C_A)$ for a gluon.

Finally we have the soft wide-angle single-logarithmic contribution
$\ln S$, which depends on the geometry of the hard three-jet system
via the dependence on dipole invariant masses $Q_{ij} =2 (p_i.p_j)$.
This structure is characteristic of soft inter-jet radiation for
three-jet systems (see e.g ref.~\cite{yupinbanzan} for a detailed
discussion). The result can be easily extended to $2 \to 2$ hard
processes as shown below.

\subsection{Radiator for hadron collisions}

One can easily generalise the results just presented to the case of
azimuthal decorrelations in hadron collisions such as at the Tevatron
and LHC. Here one is dealing with the suppression of radiation from an
ensemble of four hard partons since one has two incoming legs in
addition to the final-state dijet system.

One considers each partonic subprocess and obtains an equation similar
to eq.~\eqref{eq:intcross} involving this time two pdfs for the
incoming partons. The $b$-space radiator can once again be computed
dipole-by--dipole (i.e. for each pair of hard partons) and the results
combined after weighting by the colour factor for each dipole. It
reads:
\begin{eqnarray}
\label{eq:hh}
R_{\mathrm{out}}(\bar{b}) &=& (C_1+C_2) \frac{\alpha_s}{2 \pi}
\left(\frac{2}{3} L^2 + \frac{4}{3} L \left(-\ln 3 -4 \ln 2+3
\ln \frac{Q_{12}}{p_t} \right) \right)+ \nonumber\\
&&+ \frac{4}{3} \left(C_1 B_1+C_2 B_2 \right) \frac{\alpha_s}{2\pi}
L , \\
R_{\mathrm{in}}(\bar b) &=& \left(C_{i1}+C_{i2}\right) \frac{\alpha_s}{2
\pi}\left(2 L^2 +4L \left(-\ln 2 + \ln \frac{Q_{12}}{p_t} \right) \right) +\\
&&+ 4\left(C_{i1}B_{i1}+C_{i2}B_{i2}\right) \frac{\alpha_s}{2\pi} L,
\nonumber \\ \ln S(\bar{b},\{p\})  &=& \ln \frac{\mathrm{Tr}\left(He^{-t
\Gamma^{\dagger}/2} M e^{-t \Gamma/2}
\right)}{\mathrm{Tr}\left(HM\right)},
\end{eqnarray}
with $t = 2\alpha_s L/\pi $ for a fixed coupling, and $C_{i1}$ and
$C_{i2}$ being the colour charges for the incoming partons and $C_1$
and $C_2$ those for the outgoing jets. The structure of the result
is thus similar to that for the DIS (three hard parton) case except
the function on the last line of the above equation, characteristic
of soft wide-angle gluon radiation from an ensemble of four hard
partons~\cite{Sterman}. Here the quantity $\Gamma$ is an anomalous
dimension matrix while $H$ consists of elements $H_{ij}$
representing the product of the Born amplitude in colour channel $i$
and its complex conjugate in channel $j$. Lastly the matrix $M$
represents a normalisation arising from the colour algebra. These
matrices depend on the exact $2 \to 2$ scattering channel considered
as well as on the choice of colour basis (see e.g.
ref.~\cite{Sterman} for their explicit forms in particular bases).
In the end one sums over all channels after folding the
$\Sigma(\Delta)$ for each channel with the corresponding Born
weights to obtain the final result.

\subsection{Non-global variants}

We have seen that in the $E$-scheme, used at the Tevatron, the
observable at hand is non-global and hence the resummation differs
significantly from that detailed above. At the leading logarithmic
level there are no double logarithms arising from the final-state
jets, and so the collinear pieces proportional to the colour factors
of outgoing jets in eq.~\eqref{eq:hh} would be absent. At the level
of next-to--leading logarithmic terms, two additional pieces arise.
One is the non-global piece computed for the two-jet case in for
instance ref.~\cite{DassalNG1}, which is concerned with correlated
multiple soft emission and can only be computed in the large $N_c$
approximation. The other piece is an ``independent emission''
contribution arising purely from the jet algorithm dependence which
corrects eq.~\eqref{eq:hh} at the NLL level; in particular for the
$k_t$ clustering algorithm this factor was seen to scale as $R^3$ at
leading order~\cite{BanDasLett,DelBanDas}.  However, due to the fact
that these effects arise first at $\mathcal{O}\left(\alpha_s^2
\right)$, they become significant only in a region where the
integrand in eq.~\eqref{eq:intcross} is numerically
small~\cite{BanDasSymm}. For this reason they have a negligible
impact on the resummation and can safely be ignored. Thus in
practice we are able to provide a resummed
result~\cite{BanDasDelprep} also for the current experimental
definition at the Tevatron.

\section{Results and Discussion}

To provide a final resummed result for the $\Delta \phi$
distribution one still needs to carry out the $b$ integration in
eq.~\eqref{eq:intcross}.  In this section we describe how to produce
numerically resummed differential distributions for $\Delta$ (and
hence for $\Delta \phi$) in DIS, starting from the resummed
expression eq.~\eqref{eq:sigma-fin}, summed over the two incoming
channels $a=q,g$ (the generalisation to hadron collisions should
then be obvious since the same considerations will apply there):
\begin{equation}
\label{eq:resum-start} \sigma(\Delta) = \sum_{a=q,g} \int d{\cal
B}\>|M_{\cal B}|^2 \, \Sigma_a(\{p\},\Delta)\,.
\end{equation}
We recall that the measure $d\mathcal{B}$ contains implicitly
the acceptance cuts on
the jet momenta, which in this case coincide with the Born momenta
$\{p\} $.  The function $\Sigma_a(\{p\},\Delta)$ is the NLL resummed
distribution, which can be written as:
\begin{equation} \label{eq:Sigma-DIS}
\Sigma_a(\{p\},\Delta) = \frac{2}{\pi}
\int_0^{\infty}\!\!\frac{db}{b}\> \sin(b\Delta)
f_{a}\left(x,\mu_f^2/\bar b^2\right) e^{-R^a_\mathrm{in}(\bar b)}
e^{-R^a_\mathrm{out}(\bar b)} S(\{p\},\bar b)\,.
\end{equation}
At NLL level, the incoming and outgoing radiators $R_\mathrm{in}$
and $R_\mathrm{out}$ and the soft function $S$ are as given
previously except that these are now re-computed with a two-loop
running coupling, which is required to achieve complete NLL
accuracy.

However we now have to deal with an issue that has long plagued such
resummations in $b$-space in that some reasonable but perhaps
somewhat ad-hoc prescriptions have to be adopted to practically
evaluate the $b$ integral in eq.~\eqref{eq:sigma-fin} (see for
instance the energy-energy correlation in ref.~\cite{EEC} for
related discussions):
\begin{enumerate}
\item At large $b$ the running coupling used to evaluate $R(\bar b)$
  hits the Landau pole. We then decide to simply cut-off the
  $b$ integral (i.e set $R(\bar b) = \infty$) at $\bar b =
  \exp(1/(2\alpha_s(\mu_R)\beta_0))$ which corresponds to the Landau
  pole singularity.
\item Additionally at large $b$, the factorisation scale $\mu_f/\bar
  b$ of the parton density in eq.~(\ref{eq:Sigma-DIS}) becomes small.
  Current parameterisations of parton densities usually fail for
  factorisation scales less than $Q_0 = 1$ GeV, corresponding to the
  fact that here it is not possible to neglect the intrinsic motion of
  partons inside the proton. This treatment is beyond the scope of the
  present letter, so we freeze the parton density at $Q_0$ for
  $\bar b>\mu_f/Q_0$.
\begin{figure}
\label{fig:delphi}
\begin{center}
\epsfig{file =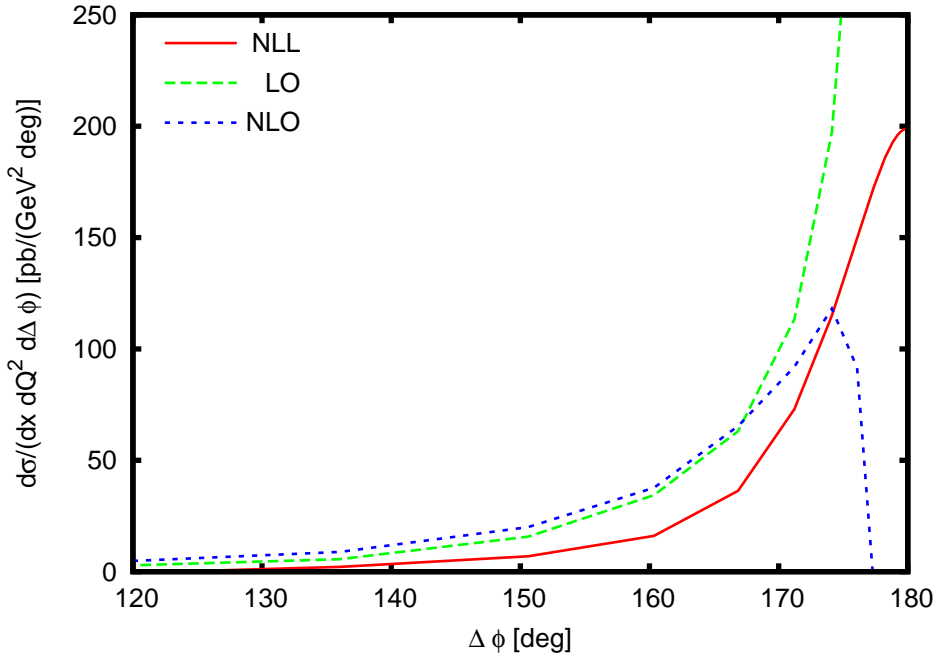}
\caption{The resummed $\Delta \phi$ distribution for dijets in DIS.
  Also shown for comparison are the leading order (LO) and
  next-to--leading order (NLO) predictions from NLOJET++~\cite{Nagy}.}
\end{center}
\end{figure}
\item At small $b$ the radiator in eq.~(\ref{eq:R-start}) ought to
vanish for $b=0$, since this point corresponds to a complete
cancelation between real and virtual correction terms. This
cancelation is not present in the pure NLL approximation we obtain
here. We have therefore set $R_\mathrm{in/out}(\bar b)=S(\bar{b})=0$
and freezed the parton density at $\mu_f$ for $\bar b < 1$. Other
modifications such as the replacement $\bar b \to \sqrt{1+\bar b^2}$
that ensure a sensible $b=0$ behaviour can also be made but we have
checked that this does not numerically alter our results presented
here.
\end{enumerate}

An important thing to notice is that an NLL resummation is strictly
valid for $1<\bar b<\mu_f/Q_0$. In particular the relative size of
the large-$b$ part of the integral, corresponding to $\bar b
>\mu_f/Q_0$, can give us an idea of the impact of intrinsic parton
transverse momentum, an area that we shall explore in more detail in
forthcoming work~\cite{BanDasDelprep}.

We plot the resummed result for the $\Delta \phi$ distribution in
fig.~\ref{fig:delphi} along with the fixed-order predictions for
dijet production in DIS with $Q^2=67$ ${\mathrm{GeV}}^2$ and
$x_B=2.86 \times 10^{-3}$. These values and other cuts on the jets
have been taken from the H1 study~\cite{Hansson} to which we would
eventually compare our results. We fixed both renormalisation and
factorisation scales to be the average transverse energy of the
jets, and used CTEQ6M parton distributions~\cite{CTEQ},
corresponding to $\alpha_s(M_Z)=0.118$. As we can see the
fixed-order predictions diverge as expected near $\Delta \phi =\pi$.
This divergence is cured by the resummation that goes to a fixed
{\it{non-zero}} value at $\Delta\phi =\pi$. Of note here is the
absence of a Sudakov peak since the Sudakov mechanism does not
dominate the $b$ integral at very small $\Delta = |\pi-\Delta\phi|.$
The dominant mechanism to obtain back-to--back jets is thus a
one-dimensional cancelation between emissions rather than a
suppression of the $k_t$ of each individual emission, leading to a
washout of the Sudakov peak as explained in detail in
ref.~\cite{BanSmyeMark}.

In order to obtain complete predictions which can be compared to
data two further developments need to be made. The first concerns
matching to fixed order which is non-trivial since it requires
information on the flavour of all the partons in the event which is
not directly available in the fixed-order codes. Here we hope to
exploit recent developments in this regard~\cite{Andrea} which have
addressed these issues in the context of hadron collider event
shapes~\cite{BanZanSal}. Secondly non-perturbative effects are
expected to play an important role in the region $\Delta \phi
\approx \pi$ where they can be expected to significantly change the
value of the distribution. Here one can apply a Gaussian smearing to
our $b$-space results as is the practice for vector boson $Q_t$
spectra~\cite{Nadolsky} as a model for non-perturbative effects
whose parameters can be constrained phenomenologically. We leave
both these developments for forthcoming work~\cite{BanDasDelprep}.

\section*{Acknowledgements}

One of us (MD) wishes to thank Gavin Salam and Lorenzo Magnea for
useful discussions concerning jets and resummation.

\end{document}